\documentclass[bib]{statapress}

\usepackage[crop,newcenter,frame]{pagedims}

\usepackage{sj}

\usepackage{epsfig}

\usepackage{stata}

\usepackage{amsmath}

\usepackage{shadow}

\sjsetissue{$vv$}{$ii$}{$mm$}{$yyyy$}

\usepackage{threeparttable} 
\usepackage{tabularx}

\usepackage[inline]{enumitem}
\usepackage{enumerate}

\usepackage{hyperref}
\hypersetup{
     colorlinks=true,
     linkcolor=black,
     filecolor=blue,
     citecolor=blue,      
     urlcolor=blue,
     }
     
\usepackage[toc,acronym,nonumberlist]{glossaries}
\usepackage[intoc, english]{nomencl} 
\makeglossaries

\usepackage{tikz,xcolor,hyperref}
\definecolor{lime}{HTML}{A6CE39}
\DeclareRobustCommand{\orcidicon}{%
	\begin{tikzpicture}
	\draw[lime, fill=lime] (0,0) 
	circle [radius=0.16] 
	node[white] {{\fontfamily{qag}\selectfont \tiny ID}};
	\draw[white, fill=white] (-0.0625,0.095) 
	circle [radius=0.007];
	\end{tikzpicture}
	\hspace{-2mm}
}

\foreach \x in {A, ..., Z}{%
	\expandafter\xdef\csname orcid\x\endcsname{\noexpand\href{https://orcid.org/\csname orcidauthor\x\endcsname}{\noexpand\orcidicon}}
}

\begin{document}

\newacronym{RUM}{RUM}{Random Utility Maximization}
\newacronym{RRM}{RRM}{Random Regret Minimization}
\newacronym{Mixed RRM}{Mixed RRM}{Mixed Random Regret Minimization}
\newacronym{SML}{SML}{Simulated Maximum Likelihood}
\newacronym{ASC}{ASC}{Alternative Specific Constants}
\newacronym{SC}{SC}{Stated Choice}


\inserttype[notag]{article}
\author{Z. Zhu, Á. A. Gutiérrez-Vargas and M. Vandebroek}{%
  Ziyue Zhu \orcidA{} \\Faculty of Sciences\\KU Leuven\\Leuven, Belgium\\ziyue.zhu16@gmail.com
  \and
  Álvaro A. Gutiérrez-Vargas \orcidB{}  \\Faculty of Economics and Business\\KU Leuven\\Leuven, Belgium\\alvaro.gutierrezvargas@kuleuven.be
  \vspace{6pt} \and 
  Martina Vandebroek \orcidC{} \\Faculty of Economics and Business\\KU Leuven\\Leuven, Belgium\\martina.vandebroek@kuleuven.be
}

\title[Fitting mixed logit random regret minimization models]{Fitting mixed logit random regret minimization models using maximum simulated likelihood}
\maketitle

\begin{abstract}
This article describes the {\tt mixrandregret} command, which extends the {\tt randregret} command introduced in \citeauthor{gutierrez2021randregret} (\citeyear{gutierrez2021randregret}, \textit{The Stata Journal} 21: 626–658) incorporating random coefficients for Random Regret Minimization models. The newly developed command {\tt mixrandregret} allows the inclusion of random coefficients in the regret function of the classical RRM model introduced in \citeauthor{chorus2010new} (\citeyear{chorus2010new}, \textit{European Journal of Transport and Infrastructure Research} 10: 181-196). The command allows the user to specify a combination of fixed and random coefficients. In addition, the user can specify normal and log-normal distributions for the random coefficients using the commands’ options. The models are fitted using simulated maximum likelihood using numerical integration to approximate the choice probabilities. 

\end{abstract}

\section{Introduction}
\cite{McFadden1974} introduced conditional logit models to explain the choice behavior of individuals and to predict market shares of products and services. The conditional logit models form the basis for the majority of discrete choice models, which assume that individuals use a decision rule based on \gls{RUM} when choosing between various alternatives. In contrast, \cite{chorus2008random} proposed an alternative decision rule known as \gls{RRM}, assuming that decision-makers aim to minimize regret when making their choices. \cite{RePEc:jae:japmet:v:15:y:2000:i:5:p:447-470} extended the random utility model by allowing the parameters to vary across individuals, leading to the so-called mixed logit model. Similarly, \cite{hensher2016random} modified the \gls{RRM} models to include random effects, which account for preference heterogeneity and allow for correlation among choices made by the same individual. 

In this article, we extend the command {\tt randregret} \citep{gutierrez2021randregret} into a mixed version called {\tt mixrandregret} which allows the inclusion of random parameters. The new command allows the user to specify normal and log-normally distributed taste parameters inside the regret function. The parameters of the distribution of the coefficients are estimated using \gls{SML}. Specifically, given that there is no closed-form solution for the choice probabilities, we approximate them using simulations. We also developed the {\tt mixrpred} post-estimation command that can predict the choice probabilities for each alternative. Additionally, the {\tt mixrbeta} post-estimation command allows estimating the individual-level parameters for each individual. We will illustrate the command's usage in examples from \cite{vanCranenburgh2018}.

\section{Classical Random Regret Models}

In contrast to the decision-making process of \gls{RUM} models, which measure the benefits of selecting a particular alternative in terms of utility, \gls{RRM} models focus on the regret resulting from not-chosen alternatives. Regret occurs when, compared to other available alternatives, the selected alternative is outperformed by the other alternatives in some of the attributes \citep{loomes1982regret}. Accordingly, \gls{RRM} models assume that the individuals intend to minimize regret when choosing among alternatives. Formally, \cite{chorus2008random} presented an initial model for random regret minimization models, and \cite{chorus2010new} revised the regret function in order to obtain a smooth likelihood function. Accordingly, he proposed \eqref{SysReg} to denote the regret of individual $n$ when choosing alternative $i$ among the $J$ possible alternatives
\begin{equation}\label{SysReg}
\textsc{R}_{in} = \sum_{j\neq i}^J \sum_{m=1}^M \ln[1+\exp\{\beta_{m} \cdot (x_{jn,m} - x_{in,m})\}] + \alpha_i,
\end{equation}

Equation \eqref{SysReg}  represents the regret that an individual (referred to by $n$) experiences when choosing alternative $i$ among $J$ alternatives (referred to by $j$ or $i$). Additionally, each alternative is described in terms of the value of $M$ attributes (referred to by $m$). Consequently, $x_{in,m}$ represents the values of attribute $m$ of alternative $i$ for individual $n$, and $\beta_{m}$ is the taste parameter of attribute $m$ for individual $n$. The parameter $\beta_{m}$ indicates that for each unit change of attribute $m$ in a non-selected alternative, regret would either increase (if $\beta_m$ is positive) or decrease (if $\beta_m$ is negative) relative to the level of the same attribute in the selected alternative. Besides, the inclusion of \gls{ASC} in the stated models is possible by simply adding them to the systematic part of the regret as $\alpha_i$. The inclusion of the \gls{ASC} serves the same purpose as in \gls{RUM} models, which is to account for omitted attributes for a particular alternative $i$. As usual, for identification purposes, we need to exclude one of the \gls{ASC} from the model specification, so we define $\boldsymbol{\alpha}=(\alpha_{i},\dots,\alpha_{J-1})$ as the vector of $J-1$ \gls{ASC} included in the model. A detailed discussion of the \gls{ASC} in the context of \gls{RRM} models see \cite{van2016robustness}. Consequently, $\textsc{R}_{in}$ describes the total systematic regret for an individual $n$ choosing alternative $i$. 




Similarly to \gls{RUM} models, we can obtain the random regret function, $\textsc{RR}_{in}$, by adding an i.i.d extreme value type I error term to the systematic regret function, $\textsc{R}_{in}$, that will account for the pure random noise and the impact of omitted attributes in the regret function: $\textsc{RR}_{in} = \textsc{R}_{in} + \varepsilon_{in}$. Mathematically, the minimization of the random regret function is equivalent to maximizing the negative function, which results in the conventional closed-form logit formula for the choice probabilities given in equation \eqref{eq:RRMProb}.
\begin{equation}\label{eq:RRMProb}
P_{in} = \frac{\exp(-{R}_{in})}{\sum_{j=1}^{J}\exp(-{R}_{jn})}.
\end{equation}

The log-likelihood function of the regret model for $N$ individuals is given by equation \eqref{eq:RRMlog}, where $\boldsymbol{\beta} = (\beta_{1},\dots,\beta_{m})$ is the vector of taste parameters and $y_{in}$ is the dummy variable that takes the value of 1 when alternative $i$ is chosen by individual $n$, and 0 otherwise.
\begin{equation}\label{eq:RRMlog}
LL(\boldsymbol{\alpha}, \boldsymbol{\beta} ) 
=
\sum_{n=1}^N\sum_{s=1}^S\sum_{i=1}^J 
y_{in}
\times
\ln\left(P_{in}\right).
\end{equation}
In the literature, there exist several extensions to the classical \gls{RRM} models \citep{chorus2014generalized,van2015new}. \cite{chorus2014generalized} proposed the generalized \gls{RRM}, which replaces the ``1'' in the regret function with a new parameter $\gamma_m$ denoting the regret-weight for attribute $m$. \cite{van2015new} incorporated a scale parameter into the \gls{RRM}, which is now referred to as $\mu$\gls{RRM}. The Pure-\gls{RRM} was proposed in the same article \citep{van2015new}, as a special case of $\mu$\gls{RRM} when $\mu$ arbitrarily small. For a review that compares the different types of \gls{RRM} models and \gls{RUM} models, see \cite{gutierrez2021randregret}. In what follows, we will focus on the classical regret function of \cite{chorus2010new}, but we will allow for the inclusion of random taste parameters as introduced by \cite{hensher2016random}. This model will be referred to as the \gls{Mixed RRM} model and takes preference heterogeneity into consideration by assuming a parametric distribution for the taste parameters. 


\section{Mixed Random Regret Minimization Models}

In this section, we describe the \gls{Mixed RRM} where we 
\begin{enumerate*}[label*=(\roman*)]
  \item allow that the taste parameters follow a parametric distribution, and\label{item_1_mixrrm}
  \item we are able to model data with panel structure. \label{item_2_mixrrm}
\end{enumerate*}Consequently, \ref{item_1_mixrrm} triggers a new sub-index to the taste parameters, $\boldsymbol{\beta}_n = (\beta_{n,1},\dots,\beta_{n,m})$, which now follow a parametric distribution $f(\boldsymbol{\beta}| \boldsymbol{\varphi})$, where $\boldsymbol{\varphi}$ are the parameters that describe the distribution\footnote{For instance, if we assume a normal distribution, $\boldsymbol{\varphi}$ would contain its mean and variance.}. Hence, $\beta_{n,m}$ is now an individual-specific taste parameter that represents the regret sensitivity of individual $n$ to changes in attribute $m$. Additionally, \ref{item_2_mixrrm} implies that multiple choice situations (referred to by $s$) are answered by the same individual, which triggers the inclusion of a new sub-index for the choice situations in our formulas. Hence, $x_{ins,m}$ will now represent the value of attribute $m$ for alternative $i$ for individual $n$ in choice situation $s$. Similarly, $y_{ins}$ is now a binary variable that takes the value of 1 when individual $n$ choose alternative $i$ in choice situation $s$ and 0 otherwise. That being said, we will define a new regret function that considers points \ref{item_1_mixrrm} and \ref{item_2_mixrrm} in equation \eqref{mixSysReg} where $R_{ins}$ describes the systematic regret for individual $n$ choosing alternative $i$ in choice situation $s$.
\begin{equation}\label{mixSysReg}
R_{ins} = \sum_{j\neq i}^J \sum_{m=1}^M \ln[1+\exp\{\beta_{n,m} \cdot (x_{jns,m} - x_{ins,m})\}] + \alpha_i,
\end{equation}
 Similarly, we add the i.i.d extreme value type I error term to the systematic regret function, and the choice probability is given by equation \eqref{eq:mixedRRMchoiceProb}.
\begin{equation}\label{eq:mixedRRMchoiceProb}
P_{ins} = \frac{\exp(-R_{ins})}{\sum_{j=1}^{J}\exp(-R_{jns})}.
\end{equation}
Additionally, the probability of the observed sequence of choices of individual $n$ (conditional on knowing $\boldsymbol{\beta}_n$) is given by equation \eqref{eq:mixedRRMProb}, which differs from equation \eqref{eq:RRMProb} in the sense that equation \eqref{eq:mixedRRMProb} consider responses from the same individual might be correlated, but responses from different individuals are treated as independent from one another. 
\begin{align}\label{eq:mixedRRMProb}
P_{n}(\boldsymbol{\alpha},\boldsymbol{\beta}) 
 =  \prod_{s=1}^{S} \prod_{j=1}^{J} \{P_{ins}\}^{y_{ins}}.
\end{align}
The unconditional choice probabilities of the observed sequence of choices are the conditional choice probabilities (see equation \ref{eq:mixedRRMProb}) integrated over the entire domain of the distribution. Consequently, the log-likelihood function of the \gls{Mixed RRM} Model in equation \eqref{eq:MIXL-LL}.
\begin{align}\label{eq:MIXL-LL}
LL(\boldsymbol{\alpha},\boldsymbol{\varphi}) 
= 
\sum_{n=1}^{N} \ln
\left[
\int_{\boldsymbol{\beta}} 
P_{n}(\boldsymbol{\alpha},\boldsymbol{\beta}) 
f(\boldsymbol{\beta}| \boldsymbol{\varphi})
d\boldsymbol{\beta}
\right]
\end{align}
Given that the integral described in equation \eqref{eq:MIXL-LL} does not have a closed-form solution, it is approximated using simulation \citep{train2009discrete}. Accordingly, we estimate the model by \gls{SML}. Hence, we maximize the simulated log-likelihood function of equation \eqref{eq:simulated_LL_MIXL} where $R$ is the number of draws and $\boldsymbol{\beta}^{r}$ is the $r$th drawn from $f(\boldsymbol{\beta}| \boldsymbol{\varphi})$. Finally, we use Halton draws to create the draws used to approximate the choice probabilities.

\begin{equation}\label{eq:simulated_LL_MIXL}
SLL(\boldsymbol{\alpha},\boldsymbol{\varphi}) 
= 
\sum_{n=1}^{N}
\ln \left\{ 
\frac{1}{R} 
\sum_{r=1}^{R} 
P_{n}(\boldsymbol{\alpha},\boldsymbol{\beta}^r)
\right\}
\end{equation}

\section{Individual-level Parameters}
After maximizing the simulated log-likelihood function to obtain estimates for $\boldsymbol{\hat{\varphi}}$ and $\boldsymbol{\hat{\alpha}}$, we can also obtain estimates for the individual-level parameters. That is to say, we can estimate the taste parameters for every individual conditional on their sequences of choices (denoted by $\boldsymbol{y}_{n}$) and the attribute levels for every alternative and choice set, denoted by $\boldsymbol{x}_{n}$, that the individual faced when making the choices. For instance, we can compute the individual-level parameter $\bar{\boldsymbol{\beta}}_n$ for every individual $n$ which corresponds to the mean of the distribution of $\boldsymbol{\beta}_n$ conditional on $\boldsymbol{y}_{n}$, $\boldsymbol{x}_{n}$, and our estimated $\hat{\boldsymbol{\varphi}}$. The expression for $\bar{\boldsymbol{\beta}}_n$ is given in equation \eqref{eq:meanb}, and its derivation can be found in \cite{train2009discrete}:

\begin{align} \label{eq:meanb}
    \boldsymbol{\bar{\beta}}_n 
     = \frac{ \int_{\boldsymbol{\beta}} \boldsymbol{\beta}
     \times
     P_n(\boldsymbol{y}_n | \boldsymbol{x}_n, \boldsymbol{\hat{\alpha}},\boldsymbol{\beta})
     f(\boldsymbol{\beta} | \hat{\boldsymbol{\varphi}}) d\boldsymbol{\beta}}{\int_{\boldsymbol{\beta}} P_n(\boldsymbol{y}_n | \boldsymbol{x}_n, \boldsymbol{\hat{\alpha}}, \boldsymbol{\beta}) f(\boldsymbol{\beta} | \hat{\boldsymbol{\varphi}}) d\boldsymbol{\beta}}.
\end{align}

Again, since there is no closed-form solution for the integrals in equation \eqref{eq:meanb}, we approximate them using simulations yielding to equation \eqref{eq:simMean}:
\begin{equation}\label{eq:simMean}
    \sthat{\boldsymbol{\beta}_{n}} = \sum^R_{r=1} \left( \frac{ \boldsymbol{\beta}^r \times P_n(\boldsymbol{y}_n | \boldsymbol{x}_n, \boldsymbol{\hat{\alpha}}, \boldsymbol{\beta}^r)}{\sum^R_{r=1} P_n(\boldsymbol{y}_n | \boldsymbol{x}_n, \boldsymbol{\hat{\alpha}}, \boldsymbol{\beta}^r)}\right),
\end{equation}
where $R$ is the number of draws and $\boldsymbol{\beta}^{r}$ is the $r$th drawn from $f(\boldsymbol{\beta}| \boldsymbol{\varphi})$. 

\section{Commands}

\subsection{mixrandregret}

\subsubsection*{Syntax}

\begin{stsyntax}
    mixrandregret
    \depvar\
    \optindepvars\
    \optif\
    \optin\
    \optweight\
    ,
    \tt id(\varname) 
    \underbar{gr}oup(\varname)
    rand(\varlist)
    \underbar{al}ernatives(\varname)
    \optional{\underbar{base}alternatives(\num)
    \underbar{nocons}tant
    \underbar{cl}uster(\varname)
    \underbar{r}obust
    ln(\num)
    nrep(\num)
    burn(\num)
    \underbar{l}evel(\num)
    {\it maximize\_options}}
\end{stsyntax}

\bigskip
\hangpara $\depvar$ equal to 1 identifies the chosen alternative, whereas a 0 indicates that the alternative was not selected. There is only one chosen alternative for each choice set.

\hangpara $fweights$, $iweights$, and $pweights$ are allowed (see \uref{weight}), but they are applied to decision-makers, not to individual observations.

\subsubsection*{Description}
\texttt{mixrandregret} estimates the mixed random regret minimization model described in \cite{hensher2016random}, which is a mixed version of the classic random regret minimization model introduced in \cite{chorus2010new}. {\tt mixrandregret} extends the {\tt randregret} command \citep{gutierrez2021randregret} and allows the user to specify normally and log-normally distributed taste parameters inside the regret function. The command uses simulated maximum likelihood for estimation \citep{train2009discrete}. 

\subsubsection*{Options}
\hangpara {\tt id(\varname)} is required and specifies a numeric identifier variable for the decision-makers.

\hangpara {\tt \underbar{gr}oup(\varname)} is required and specifies a numeric identifier variable for the choice occasions.

\hangpara {\tt rand(\varlist)} is required and specifies the independent variables whose coefficients are random. The random coefficients can be specified to be normally or log-normally distributed (see the {\tt ln()} option). The variables immediately following the dependent variable in the syntax are specified to have fixed coefficients.

\hangpara {\tt \underbar{alt}ernatives(\varname)} is required to identify the alternatives available for each case.

\hangpara {\tt \underbar{base}alternatives(\num)} sets base Alternative Specific Constants (ASC) if ASC is not suppressed.

\hangpara {\tt \underbar{nocons}tant} suppress the ASC.

\hangpara {\tt \underbar{cl}uster(\varname), \underbar{r}obust} see \uref{estimation}. The cluster variable must be numeric.

\hangpara {\tt ln(\num)} specifies that the last {\tt \#} variables in {\tt rand()} have log-normally rather than normally distributed coefficients. The default is {\tt ln(0)}.

\hangpara {\tt nrep(\num)} specifies the number of Halton draws used for the simulation. The default is {\tt nrep(50)}.

\hangpara {\tt burn(\num)} specifies the number of initial elements to be dropped when creating the Halton sequences. The default is {\tt burn(15)}. Specifying this option helps reduce the correlation between the sequences in each dimension.

\hangpara {\tt \underbar{l}evel(\num)} set the confidence level. The default is {\tt level(95)}.

\hangpara {\tt {\it maximize\_options}} {\tt \underbar{dif}ficult, \underbar{tech}nique({\it algorithm\_spec}), \underbar{iter}ate(\#), \underbar{tr}ace, \underbar{grad}ient, showstep, \underbar{hess}ian, \underbar{tol}erance(\#), \underbar{ltol}erance(\#), \underbar{gtol}erance(\#), \underbar{nrtol}erance(\#), from({\it init\_specs})}; see \uref{maximize}.

\subsection{mixrpred}
\subsubsection*{Syntax}
\begin{stsyntax}
    mixrpred
    \newvarname\
    \optif\
    \optin\
    \optional{,
    proba
    nrep(\num)
    burn(\num)}
\end{stsyntax}

\subsubsection*{Description}
Following {\tt mixrandregret}, {\tt mixrpred} can be used to obtain the predicted probabilities by specifying the option {\tt proba}.

\subsubsection*{Options}
\hangpara {\tt proba} calculate the choice probability for each alternative for each choice situation; the default option.

\hangpara {\tt nrep(\num)} specifies the number of Halton draws used for the simulation. The default is {\tt nrep(50)}.

\hangpara {\tt burn(\num)} specifies the number of initial elements to be dropped when creating the Halton sequences. The default is {\tt burn(15)}. Specifying this option helps reduce the correlation between the sequences in each dimension.

\subsection{mixrbeta}
\subsubsection*{Syntax}
\begin{stsyntax}
    mixrbeta
    \varlist\
    \optif\
    \optin\
    ,
   \tt \underbar{sav}ing({\it filename})
    \optional{,
    plot
    nrep(\num)
    burn(\num)
    replace}
\end{stsyntax}

\subsubsection*{Description}
{\tt mixrbeta} can be used after {\tt mixrandregret} to calculate individual-level parameters corresponding to the variables in the specified {\it varname} using equation \eqref{eq:simMean}. The individual-level parameters are stored in a user-specified data file.

\subsubsection*{Options}
\hangpara {\tt saving({\it filename})} saves individual-level parameters to {\it filename}.

\hangpara {\tt plot} create the plots of the distribution of individual-level parameters conditional on the estimates of {\tt mixrandregret} for individual-level parameters for each individual.

\hangpara {\tt nrep(\num)} specifies the number of Halton draws used for the simulation. The default is {\tt nrep(50)}.

\hangpara {\tt burn(\num)} specifies the number of initial sequence elements to be dropped when creating the Halton sequences. The default is {\tt burn(15)}. Specifying this option helps reduce the correlation between the sequences in each dimension.

\hangpara {\tt replace} overwrites {\it filename}.

\section{Examples}
To show how we can fit \gls{Mixed RRM} Models using {\tt mixrandregret}, we use data from \cite{vanCranenburgh2018} on a \gls{SC} experiment\footnote{You can download the dataset from 4TU ResearchData: \url{https://data.4tu.nl/articles/dataset/Small\_value-of-time\_experiment\_Netherlands/12681650}}. These data are collected to analyse the impact of the different decision rules on the statistical efficiency of the design \citep{van2016robustness}. The participants answered 10 choice situations where they chose from three unlabelled route alternatives with two generic attributes: travel cost and travel time. The following variables are used in our illustration:

\begin{itemize}
    \item \texttt{altern}: identify the alternative faced by the user (sub-index {\it i} or {\it j}).
    \item \texttt{choice}: whether the alternative was chosen by the individual (dummy, 1 if chosen).
    \item \texttt{id}: ID of the individual.
    \item \texttt{cs}: ID of the choice situation faced by the individual.
    \item \texttt{tt}: total travel time of the alternative in minutes.
    \item \texttt{tc}: total travel cost of the alternative in euros.
\end{itemize}

We follow the data setup in {\tt randregret} (see \uref{randregret}), and the setup for {\tt mixrandregret} is identical to that required by {\tt mixlogit} (see \uref{mixlogit}), which is the panel representation in terms of individual-alternative. The data set is loaded from the server to Stata directly as illustrated below. We keep the variables of interest and list the first 3 observations. The data loaded are in wide format as each row corresponds to a choice situation. 

\begin{stlog}
. scalar server = "https://data.4tu.nl/ndownloader/"
{\smallskip}
. scalar doi = "files/24015353"
{\smallskip}
. import delimited "`=server + doi'",clear
(encoding automatically selected: ISO-8859-1)
(29 vars, 1,060 obs)
{\smallskip}
. keep obs id tt1 tc1 tt2 tc2 tt3 tc3 choice 
{\smallskip}
. list obs id tt1 tc1 tt2 tc2 tt3 tc3 choice in 1/3,sepby(obs)
{\smallskip}
     {\TLC}\HLI{55}{\TRC}
     {\VBAR} obs   id   tt1   tc1   tt2   tc2   tt3   tc3   choice {\VBAR}
     {\LFTT}\HLI{55}{\RGTT}
  1. {\VBAR}   1    1    23     6    27     4    35     3        3 {\VBAR}
     {\LFTT}\HLI{55}{\RGTT}
  2. {\VBAR}   2    1    27     5    35     4    23     6        2 {\VBAR}
     {\LFTT}\HLI{55}{\RGTT}
  3. {\VBAR}   3    1    35     3    23     5    31     4        1 {\VBAR}
     {\BLC}\HLI{55}{\BRC}
\nullskip
\end{stlog}

Following the data manipulation in \cite{gutierrez2021randregret}, we transform the data set using the {\tt reshape} command and present the data in long format below. We list the first 12 rows, and each row now corresponds to an alternative. The dependent variable {\tt choice} is 1 for the chosen alternative in each choice situation, and 0 otherwise. {\tt altern} identifies the alternatives in a choice situation; {\tt cs} identifies the choice situation faced by the individual; and {\tt id} identifies the individual. Furthermore, {\tt total\_time} and {\tt total\_cost}  are obtained from the {\tt tt} and {\tt tc} variables.

\begin{stlog}
. rename (choice)  (choice_w)
. qui reshape long tt tc, i(obs) j(altern)
. generate choice = 0
. replace  choice = 1 if  choice_w==altern  
. label define alt_label 1 "First" 2 "Second" 3 "Third" 
. label values altern alt_label
. gen cs  = obs
. gen total_time  = tt
. gen total_cost  = tc
. list id cs altern total_time total_cost choice in 1/12, sepby(cs) ab(10) noo
{\smallskip}
  {\TLC}\HLI{53}{\TRC}
  {\VBAR} id   cs   altern   total_time   total_cost   choice {\VBAR}
  {\LFTT}\HLI{53}{\RGTT}
  {\VBAR}  1    1    First           23            6        0 {\VBAR}
  {\VBAR}  1    1   Second           27            4        0 {\VBAR}
  {\VBAR}  1    1    Third           35            3        1 {\VBAR}
  {\LFTT}\HLI{53}{\RGTT}
  {\VBAR}  1    2    First           27            5        0 {\VBAR}
  {\VBAR}  1    2   Second           35            4        1 {\VBAR}
  {\VBAR}  1    2    Third           23            6        0 {\VBAR}
  {\LFTT}\HLI{53}{\RGTT}
  {\VBAR}  1    3    First           35            3        1 {\VBAR}
  {\VBAR}  1    3   Second           23            5        0 {\VBAR}
  {\VBAR}  1    3    Third           31            4        0 {\VBAR}
  {\LFTT}\HLI{53}{\RGTT}
  {\VBAR}  1    4    First           27            4        0 {\VBAR}
  {\VBAR}  1    4   Second           23            5        0 {\VBAR}
  {\VBAR}  1    4    Third           35            3        1 {\VBAR}
  {\BLC}\HLI{53}{\BRC}
\nullskip
\end{stlog}

We begin by fitting a classical \gls{RRM} Model using the {\tt randregret} command to obtain  reasonable starting values for {\tt mixrandregret}. We also declare {\tt noncons} suppressing the \gls{ASC} given that alternatives are non-labeled in the survey. If we have labeled data, we can specify the base alternative by declaring {\tt base()} option. As we have repeated choices from a given individual, the standard errors are corrected by specifying {\tt cluster(id)}. As expected, both parameter estimates are negative and highly significant, suggesting that regret decreases as the level of travel time or travel cost increases in a non-chosen alternative compared with the same attribute level in the chosen one. The coefficients are saved in {\tt init\_mix\_rrm} for later use as initial values for {\tt mixrandregret}.

\begin{stlog}
. randregret choice total_time total_cost, group(cs) alternatives(altern) ///
> rrmfn(classic) nocons cluster(id)
\HLI{78}
Fitting Classic RRM Model 
\HLI{78}
{\smallskip}
initial:       log likelihood =  -1164.529
alternative:   log likelihood = -1156.5784
rescale:       log likelihood =   -1121.29
Iteration 0:   log likelihood =   -1121.29  
Iteration 1:   log likelihood = -1118.4843  
Iteration 2:   log likelihood = -1118.4784  
Iteration 3:   log likelihood = -1118.4784  
{\smallskip}
RRM: Classic Random Regret Minimization Model
{\smallskip}
Case ID variable: cs                           Number of cases    =       1060
Alternative variable: altern                   Number of obs      =       3180
                                               Wald chi2(2)       =      40.41
Log likelihood = -1118.4784                    Prob > chi2        =     0.0000
                                 (Std. Err. adjusted for   106 clusters in id)
\HLI{13}{\TOPT}\HLI{64}
             {\VBAR}               Robust
      choice {\VBAR} Coefficient  std. err.      z    P>|z|     [95\% conf. interval]
\HLI{13}{\PLUS}\HLI{64}
RRM          {\VBAR}
  total_time {\VBAR}   -.102813   .0182526    -5.63   0.000    -.1385874   -.0670386
  total_cost {\VBAR}   -.417101    .068059    -6.13   0.000    -.5504943   -.2837078
\HLI{13}{\BOTT}\HLI{64}
{\smallskip}
. matrix b_rrm = e(b)
{\smallskip}
. matrix zero = J(1,1,0.01)
{\smallskip}
. matrix init_mix_rrm = b_rrm, zero
{\smallskip}
. matrix li init_mix_rrm 
{\smallskip}
init_mix_rrm[1,3]
           RRM:        RRM:            
    total_time  total_cost          c1
y1    -.102813  -.41710104         .01
{\smallskip}
\nullskip
\end{stlog}

We then fit a \gls{Mixed RRM} Model in which the coefficient for {\tt total\_cost} is fixed, but the coefficient for {\tt total\_time} is normally distributed. We use the option {\tt from()} in {\tt mixrandregret} to initialize the optimization routine using the values saved in {\tt init\_mix\_rrm} as the starting point for the mean for the {\tt total\_time} parameter. We estimated the model using 500 Halton draws to approximate the choice probabilities of equation \eqref{eq:simulated_LL_MIXL}. Additionally, we clustered our standard errors at the individual level using {\tt cluster(id)}.

\begin{stlog}
. mixrandregret choice total_cost, group(cs) alter(altern) rand(total_time) ///
> id(id) nocons cluster(id) nrep(500) from(init_mix_rrm) tech(bhhh)
{\smallskip}
Iteration 0:   log likelihood = -2850.0956  
Iteration 1:   log likelihood =  -2169.409  
Iteration 2:   log likelihood = -861.11253  
Iteration 3:   log likelihood = -771.96998  
Iteration 4:   log likelihood = -771.20333  
Iteration 5:   log likelihood = -771.09059  
Iteration 6:   log likelihood =  -771.0649  
Iteration 7:   log likelihood = -771.05912  
Iteration 8:   log likelihood = -771.05774  
Iteration 9:   log likelihood = -771.05741  
Iteration 10:  log likelihood = -771.05733  
Iteration 11:  log likelihood = -771.05731  
{\smallskip}
Case ID variable: cs                           Number of cases    =       1060
Alternative variable: altern                   
Random variable(s): total_time                 
{\smallskip}
                                 (Std. Err. adjusted for  106 clusters in id)
{\smallskip}
Mixed random regret model                               Number of obs =  3,180
                                                        Wald chi2(2)  = 606.11
Log likelihood = -771.05731                             Prob > chi2   = 0.0000
{\smallskip}
\HLI{13}{\TOPT}\HLI{64}
             {\VBAR}                 OPG
      choice {\VBAR} Coefficient  std. err.      z    P>|z|     [95\% conf. interval]
\HLI{13}{\PLUS}\HLI{64}
Mean         {\VBAR}
  total_cost {\VBAR}  -1.102136   .0449727   -24.51   0.000    -1.190281   -1.013991
  total_time {\VBAR}  -.3580736   .0581449    -6.16   0.000    -.4720355   -.2441117
\HLI{13}{\PLUS}\HLI{64}
SD           {\VBAR}
  total_time {\VBAR}   .5068268    .041366    12.25   0.000      .425751    .5879027
\HLI{13}{\BOTT}\HLI{64}
{\smallskip}
The sign of the estimated standard deviations is irrelevant: interpret them as
being positive
{\smallskip}
. matrix b_mixrrm = e(b)
{\smallskip}
\nullskip
\end{stlog}

\begin{figure}[h!]
    \centering
    \includegraphics[scale=1]{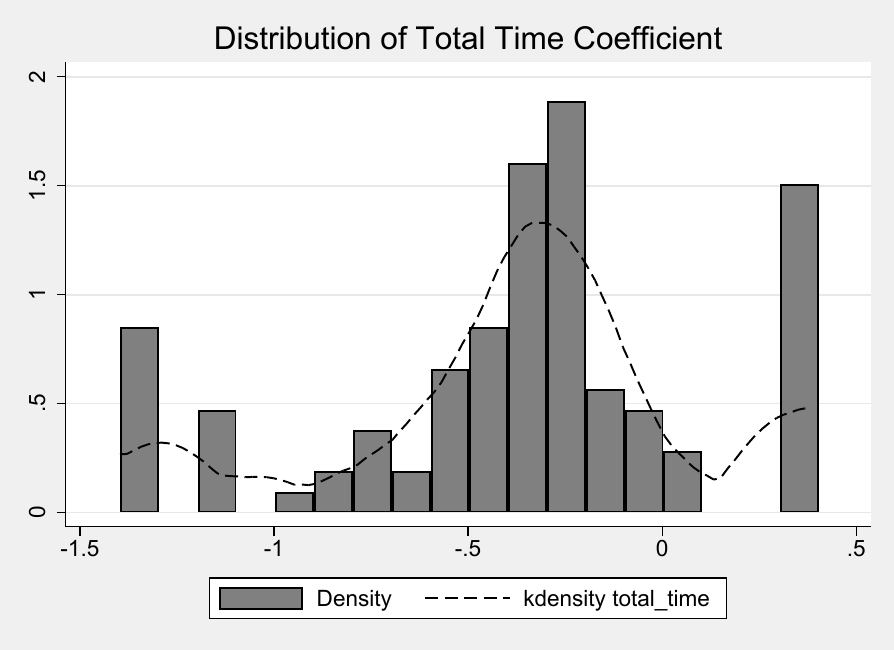}
    \caption{Distribution of Total Time Coefficient (Normal)}
    \label{fig:dis1}
\end{figure}

On average, the regret decreases as the total travel time increases in a non-chosen alternative, compared to the same level of travel time in the chosen alternative. The interpretation is similar for the total travel cost attribute. Additionally, we observe that there is significant regret heterogeneity for total travel time, given that the standard deviation parameter for total travel time is statistically different from zero. Furthermore, after the estimation of the \gls{Mixed RRM} Model, we can compute individual-level parameters using {\tt mixrbeta}. In the code below, we use equation \eqref{eq:simMean} to approximate the value for the regret coefficient for each individual using 500 Halton draws. Additionally, {\tt mixrbeta} creates a new data set with one observation per individual ({\tt id}) and its corresponding parameter estimates. Subsequently, we also display the estimates for the first five individuals in the sample, where we observe that some of them have a positive coefficient for the {\tt total\_time} attribute. Besides, we plot the individual level parameters for {\tt total\_time} in Figure \ref{fig:dis1} for all the individuals in the sample and observe that there are individuals with positive estimates for the {\tt total\_time} coefficient, which is counter-intuitive.

\begin{stlog}
{\smallskip}
. mixrbeta total_time, nrep(500) replace saving("${\lbr}graphs_route{\rbr}\\mixRRM_normal_idl") 
{\smallskip}
. use "${\lbr}graphs_route{\rbr}\\mixRRM_normal_idl", replace
{\smallskip}
. list id total_time in 1/5 
{\smallskip}
     {\TLC}\HLI{17}{\TRC}
     {\VBAR} id   total_time {\VBAR}
     {\LFTT}\HLI{17}{\RGTT}
  1. {\VBAR}  1    .37640482 {\VBAR}
  2. {\VBAR}  2   -.05517462 {\VBAR}
  3. {\VBAR}  3    .37672848 {\VBAR}
  4. {\VBAR}  4    .38495822 {\VBAR}
  5. {\VBAR}  5    .37607978 {\VBAR}
     {\BLC}\HLI{17}{\BRC}
\nullskip
\end{stlog}

One solution to obtain non-positive estimates for the {\tt total\_time} coefficient is to use a bounded distribution. When using {\tt mixrandregret}, we can specify that a coefficient is log-normally distributed for this purpose. In our case, since we want a non-positive distribution for the {\tt total\_time} coefficient, we have to multiply the  {\tt total\_time} attribute for -1 to ensure that it is non-positive. To this end we create the new variable {\tt ntt}, which corresponds to the negative of {\tt total\_time}.

\begin{stlog}
. gen ntt = -1 * total_time
{\smallskip}
. mixrandregret choice total_cost, group(cs) alt(altern) rand(ntt) ln(1) id(id) ///
> nocons cluster(id) nrep(500) tech(bhhh) from(b_mixrrm)
{\smallskip}
Iteration 0:   log likelihood = -994.35461  
Iteration 1:   log likelihood = -858.23241  
Iteration 2:   log likelihood =  -798.4694  
Iteration 3:   log likelihood = -785.66872  
Iteration 4:   log likelihood = -785.30817  
Iteration 5:   log likelihood = -785.27945  
Iteration 6:   log likelihood = -785.27728  
Iteration 7:   log likelihood = -785.27686  
Iteration 8:   log likelihood = -785.27675  
Iteration 9:   log likelihood = -785.27672  
Iteration 10:  log likelihood = -785.27671  
{\smallskip}
Case ID variable: cs                           Number of cases    =       1060
Alternative variable: altern                   
Random variable(s): ntt                        
{\smallskip}
                                 (Std. Err. adjusted for  106 clusters in id)
{\smallskip}
Mixed random regret model                              Number of obs =   3,180
                                                       Wald chi2(2)  = 1230.55
Log likelihood = -785.27671                            Prob > chi2   =  0.0000
{\smallskip}
\HLI{13}{\TOPT}\HLI{64}
             {\VBAR}                 OPG
      choice {\VBAR} Coefficient  std. err.      z    P>|z|     [95\% conf. interval]
\HLI{13}{\PLUS}\HLI{64}
Mean         {\VBAR}
  total_cost {\VBAR}  -1.217682   .0442047   -27.55   0.000    -1.304321   -1.131042
         ntt {\VBAR}  -1.312285   .1562202    -8.40   0.000    -1.618471   -1.006099
\HLI{13}{\PLUS}\HLI{64}
SD           {\VBAR}
         ntt {\VBAR}   1.363632   .1185994    11.50   0.000     1.131181    1.596082
\HLI{13}{\BOTT}\HLI{64}
{\smallskip}
The sign of the estimated standard deviations is irrelevant: interpret them as
being positive
\nullskip
\end{stlog}

The estimated {\tt ntt} parameters are the mean and standard deviation of the natural logarithm of the coefficient, and we can transform them back to the estimates of the coefficients themselves. The median of the coefficient is given by $\exp(b_{ntt})$, the mean is given by $\exp(b_{ntt}+s_{ntt}^2/2)$, and the standard deviation is given by $\exp(b_{ntt}+s_{ntt}^2/2) \times \sqrt{\exp(s_{ntt}^2)-1}$ \citep{train2009discrete}. The sign change prior to the estimation is reversed by multiplying the estimates by -1.

\begin{stlog}
. nlcom (mean_time: -1*exp([Mean]_b[ntt]+0.5*[SD]_b[ntt]{\caret}2))
>       (med_time: -1*exp([Mean]_b[ntt])) 
>       (sd_time : exp([Mean]_b[ntt]+0.5*[SD]_b[ntt]{\caret}2)
>                  *sqrt(exp([SD]_b[ntt]{\caret}2)-1))
{\smallskip}
   mean_time: -1*exp([Mean]_b[ntt]+0.5*[SD]_b[ntt]{\caret}2)
    med_time: -1*exp([Mean]_b[ntt])
     sd_time: exp([Mean]_b[ntt]+0.5*[SD]_b[ntt]{\caret}2)*sqrt(exp([SD]_b[ntt]{\caret}2)-1)
{\smallskip}
\HLI{13}{\TOPT}\HLI{64}
      choice {\VBAR} Coefficient  Std. err.      z    P>|z|     [95\% conf. interval]
\HLI{13}{\PLUS}\HLI{64}
   mean_time {\VBAR}   -.682127   .1587961    -4.30   0.000    -.9933616   -.3708923
    med_time {\VBAR}  -.2692041   .0420551    -6.40   0.000    -.3516307   -.1867776
     sd_time {\VBAR}   1.588122   .6295756     2.52   0.012     .3541763    2.822067
\HLI{13}{\BOTT}\HLI{64}
\nullskip
\end{stlog}

\begin{figure}[h!]
    \centering
    \includegraphics[scale=1]{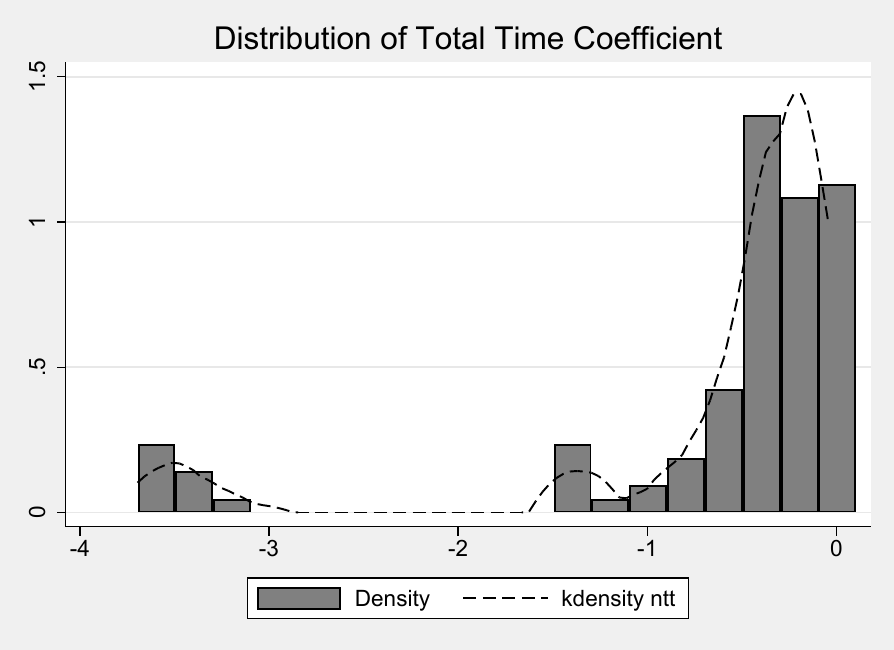}
    \caption{Distribution of Total Time Coefficient (Log-normal)}
    \label{fig:dis2}
\end{figure}

Again, we calculate individual-level parameters. As we can observe in the listed data and distribution presented in Figure \ref{fig:dis2}, all individual-level parameters are now negative as we expected.
\begin{stlog}
. mixrbeta ntt, nrep(500) replace saving("${\lbr}graphs_route{\rbr}\\mixRRM_ln_idl") 
{\smallskip}
. use "${\lbr}graphs_route{\rbr}\\mixRRM_ln_idl" , replace
{\smallskip}
. replace ntt = -1 * ntt /*reverse sign for graph*/
(106 real changes made)
{\smallskip}
. list id  ntt in 1/5 
{\smallskip}
     {\TLC}\HLI{17}{\TRC}
     {\VBAR} id          ntt {\VBAR}
     {\LFTT}\HLI{17}{\RGTT}
  1. {\VBAR}  1   -.04032598 {\VBAR}
  2. {\VBAR}  2   -.08142616 {\VBAR}
  3. {\VBAR}  3   -.04047817 {\VBAR}
  4. {\VBAR}  4   -.04110615 {\VBAR}
  5. {\VBAR}  5   -.04025335 {\VBAR}
     {\BLC}\HLI{17}{\BRC}
\nullskip
\end{stlog}

We can also generate predictions after running {\tt mixrandregret} using {\tt mixrpred}. To illustrate this command, we rerun the models using {\tt mixrandregret} with normally distributed random coefficients, suppressing the output using the {\tt quietly} command (see \uref{quietly}). Then, using the option {\tt proba}, we generate the {\tt pred\_p} variable containing the predicted probability for each alternative. The code and output are listed below.

\begin{stlog}
. qui mixrandregret choice total_cost, group(cs) alter(altern) rand(total_time) ///
> id(id) nocons cluster(id) nrep(500) from(init_mix_rrm) tech(bhhh)
{\smallskip}
. mixrpred pred_p, proba nrep(500)
{\smallskip}
{\smallskip}
. list id cs altern total_time total_cost choice pred_p in 151/162, sepby(cs) ab(10) noo
{\smallskip}
  {\TLC}\HLI{64}{\TRC}
  {\VBAR} id   cs   altern   total_time   total_cost   choice     pred_p {\VBAR}
  {\LFTT}\HLI{64}{\RGTT}
  {\VBAR}  6   51    First           23            6        0   .1516009 {\VBAR}
  {\VBAR}  6   51   Second           27            4        1   .5547292 {\VBAR}
  {\VBAR}  6   51    Third           35            3        0   .2936699 {\VBAR}
  {\LFTT}\HLI{64}{\RGTT}
  {\VBAR}  6   52    First           27            5        0   .3153724 {\VBAR}
  {\VBAR}  6   52   Second           35            4        1    .291449 {\VBAR}
  {\VBAR}  6   52    Third           23            6        0   .3931786 {\VBAR}
  {\LFTT}\HLI{64}{\RGTT}
  {\VBAR}  6   53    First           35            3        0   .3134595 {\VBAR}
  {\VBAR}  6   53   Second           23            5        1   .5523607 {\VBAR}
  {\VBAR}  6   53    Third           31            4        0   .1341798 {\VBAR}
  {\LFTT}\HLI{64}{\RGTT}
  {\VBAR}  6   54    First           27            4        0   .3153724 {\VBAR}
  {\VBAR}  6   54   Second           23            5        1   .3931786 {\VBAR}
  {\VBAR}  6   54    Third           35            3        0    .291449 {\VBAR}
  {\BLC}\HLI{64}{\BRC}
\nullskip
\end{stlog}

Additionally, {\tt mixrandregret} also allows for the inclusion of \gls{ASC} if users have labeled data. Although the data set is unlabeled in this example, we treat it as a labeled one in that each alternative represents a distinct category. We run the model including {\tt basealternative(1)} option, which specify that the first alternative is the reference group for \gls{ASC}.

\begin{stlog}
. mixrandregret choice total_cost, group(cs) alt(altern) rand(total_time) id(id) ///
> basealternative(1) cluster(id) nrep(500) tech(bhhh)
{\smallskip}
Iteration 0:   log likelihood =  -1164.529  
Iteration 1:   log likelihood = -812.87881  
Iteration 2:   log likelihood = -773.05839  
Iteration 3:   log likelihood =  -769.1873  
Iteration 4:   log likelihood = -768.22193  
Iteration 5:   log likelihood = -767.97262  
Iteration 6:   log likelihood = -767.90237  
Iteration 7:   log likelihood =  -767.8867  
Iteration 8:   log likelihood = -767.88268  
Iteration 9:   log likelihood = -767.88165  
Iteration 10:  log likelihood = -767.88138  
Iteration 11:  log likelihood = -767.88131  
Iteration 12:  log likelihood = -767.88129  
{\smallskip}
Case ID variable: cs                           Number of cases    =       1060
Alternative variable: altern                   
Random variable(s): total_time                 
{\smallskip}
                                 (Std. Err. adjusted for  106 clusters in id)
{\smallskip}
Mixed random regret model                               Number of obs =  3,180
                                                        Wald chi2(2)  = 465.50
Log likelihood = -767.88129                             Prob > chi2   = 0.0000
{\smallskip}
\HLI{13}{\TOPT}\HLI{64}
             {\VBAR}                 OPG
      choice {\VBAR} Coefficient  std. err.      z    P>|z|     [95\% conf. interval]
\HLI{13}{\PLUS}\HLI{64}
Mean         {\VBAR}
  total_cost {\VBAR}   -1.06784   .0498243   -21.43   0.000    -1.165494   -.9701866
  total_time {\VBAR}  -.3455217   .0594409    -5.81   0.000    -.4620237   -.2290197
\HLI{13}{\PLUS}\HLI{64}
SD           {\VBAR}
  total_time {\VBAR}  -.5095087   .0420965   -12.10   0.000    -.5920163   -.4270012
\HLI{13}{\PLUS}\HLI{64}
ASC          {\VBAR}
       ASC_2 {\VBAR}   .0064798   .0510223     0.13   0.899    -.0935221    .1064816
       ASC_3 {\VBAR}    .136445   .0605786     2.25   0.024     .0177131    .2551768
\HLI{13}{\BOTT}\HLI{64}
{\smallskip}
The sign of the estimated standard deviations is irrelevant: interpret them as
being positive

\nullskip
\end{stlog}

\section{Conclusions}
This article presents the command {\tt mixrandrgret} to fit Random Regret Minimization models with random parameters. We also developed the post-estimation command \texttt{mixrpred} for predicting the estimated probabilities. Additionally, the \texttt{mixrbeta} post-estimation command allows the user to estimate individual-level parameters for the random coefficients included in the model. The commands' usage and options are illustrated using discrete choice data from \cite{vanCranenburgh2018}. 

\section{Acknowledgments}
We thank Michel Meulders, Jan De Spiegeleer, and the participants from the 2022 London Stata Conference for their helpful comments and constructive suggestions. Additionally, substantial portions of our programs were inspired by the book {\it Maximum Likelihood Estimation with Stata, Fourth Edition} by Willian Gould, Jeffrey Pitblado, and Brian Poi (2010). Finally, many of the previous checks to the data and the construction of the log-likelihood functions were greatly inspired by the {\tt randregret} \citep{gutierrez2021randregret} and {\tt mixlogit} \citep{hole2007fitting} commands.

\section{Funding}
This work was produced while \'Alvaro A. Guti\'errez-Vargas was a PhD student at the Research Centre for Operations Research and Statistics (ORSTAT) at KU Leuven funded by Bijzonder Onderzoeksfonds KU Leuven (Special Research Fund KU Leuven).

\section{Conflict of interest}

Ziyue Zhu, \'Alvaro A. Guti\'errez-Vargas, and Martina Vandebroek declare no conflicts of interest.

\section{Contribution}

Ziyue Zhu and \'Alvaro A. Guti\'errez-Vargas contributed equally to the article by developing the command and drafting the article. Martina Vandebroek critically commented on both the article and the command’s functionality.

\bibliographystyle{sj}
\bibliography{references.bib}

\ifnum 13=1 \def\bibname{Reference}
\else \def\bibname{References} \fi
\begin{thebibliography}{13}
\expandafter\ifx\csname natexlab\endcsname\relax\def\natexlab#1{#1}\fi
\expandafter\ifx\csname url\endcsname\relax
  \def\url#1{\texttt{#1}}\fi
\expandafter\ifx\csname urlprefix\endcsname\relax\def\urlprefix{URL }\fi

\bibitem[{Chorus(2010)}]{chorus2010new}
Chorus, C.~G. 2010.
\newblock A new model of random regret minimization.
\newblock \emph{European Journal of Transport and Infrastructure Research}
  10(2).

\bibitem[{Chorus(2014)}]{chorus2014generalized}
\mbox{\vrule width30.25006ptheight2.62222ptdepth-2.25222pt}. 2014.
\newblock A generalized random regret minimization model.
\newblock \emph{Transportation research part B: Methodological} 68: 224--238.

\bibitem[{Chorus et~al.\@(2008)Chorus, Arentze, and
  Timmermans}]{chorus2008random}
Chorus, C.~G., T.~A. Arentze, and H.~J. Timmermans. 2008.
\newblock A random regret-minimization model of travel choice.
\newblock \emph{Transportation Research Part B: Methodological} 42(1): 1--18.

\bibitem[{van Cranenburgh and Chorus(2018)}]{vanCranenburgh2018}
van Cranenburgh, S., and C.~Chorus. 2018.
\newblock Small value-of-time experiment, Netherlands [Data set].
\newblock \emph{TU Delft - 4TU.ResearchData} .

\bibitem[{van Cranenburgh et~al.\@(2015)van Cranenburgh, Guevara, and
  Chorus}]{van2015new}
van Cranenburgh, S., C.~A. Guevara, and C.~G. Chorus. 2015.
\newblock New insights on random regret minimization models.
\newblock \emph{Transportation Research Part A: Policy and Practice} 74:
  91--109.

\bibitem[{Guti{\'e}rrez-Vargas et~al.\@(2021)Guti{\'e}rrez-Vargas, Meulders,
  and Vandebroek}]{gutierrez2021randregret}
Guti{\'e}rrez-Vargas, {\'A}.~A., M.~Meulders, and M.~Vandebroek. 2021.
\newblock randregret: A command for fitting random regret minimization models
  using Stata.
\newblock \emph{The Stata Journal} 21(3): 626--658.

\bibitem[{Hensher et~al.\@(2016)Hensher, Greene, and Ho}]{hensher2016random}
Hensher, D.~A., W.~H. Greene, and C.~Q. Ho. 2016.
\newblock Random regret minimization and random utility maximization in the
  presence of preference heterogeneity: an empirical contrast.
\newblock \emph{Journal of Transportation Engineering} 142(4): 1--10.

\bibitem[{Hole(2007)}]{hole2007fitting}
Hole, A.~R. 2007.
\newblock Fitting mixed logit models by using maximum simulated likelihood.
\newblock \emph{The stata journal} 7(3): 388--401.

\bibitem[{Loomes and Sugden(1982)}]{loomes1982regret}
Loomes, G., and R.~Sugden. 1982.
\newblock Regret theory: An alternative theory of rational choice under
  uncertainty.
\newblock \emph{The economic journal} 92(368): 805--824.

\bibitem[{McFadden(1974)}]{McFadden1974}
McFadden, D. 1974.
\newblock Conditional logit analysis of qualitative choice behavior.
\newblock \emph{In: Zarembka, P., Ed., Frontiers in Econometrics,}  105--142.

\bibitem[{McFadden and
  Train(2000)}]{RePEc:jae:japmet:v:15:y:2000:i:5:p:447-470}
McFadden, D., and K.~Train. 2000.
\newblock {Mixed MNL models for discrete response}.
\newblock \emph{Journal of Applied Econometrics} 15(5): 447--470.

\bibitem[{Train(2009)}]{train2009discrete}
Train, K.~E. 2009.
\newblock \emph{Discrete choice methods with simulation}.
\newblock Cambridge university press.

\bibitem[{Van~Cranenburgh and Prato(2016)}]{van2016robustness}
Van~Cranenburgh, S., and C.~G. Prato. 2016.
\newblock On the robustness of random regret minimization modelling outcomes
  towards omitted attributes.
\newblock \emph{Journal of choice modelling} 18: 51--70.

\end{thebibliography}

\begin{aboutauthors}
Ziyue Zhu is a master student of statistics and data science at KU Leuven in Belgium. She earned a Bachelor of Economics from Wuhan University and a Master of Economics from Barcelona School of Economics.

Álvaro A. Gutiérrez-Vargas is a PhD student at the Research Centre of Operation Research and Statistics (ORSTAT) at KU Leuven in Belgium. He earned a Bachelor of Science in economics from the University of Chile. His research interests are mainly methodological and focused on computational statistics, machine learning, and discrete choice models. He has been published in The Stata Journal and Journal of Choice Modelling.

Martina Vandebroek is a full professor at the Faculty of Economics and Business at KU Leuven in Belgium. She earned a PhD in actuarial sciences from KU Leuven. She is interested in the design of experiments, discrete choice experiments, and multivariate statistics. She has been published in Transportation Research B, Journal of Choice Modelling, Marketing Science, and Journal of Statistical Software, among other journals.
\end{aboutauthors}

\clearpage
\end{document}